\documentstyle[11pt,epsfig]{article}
\author {M. Neek-Amal$^1$ and  F. M. Peeters$^2$ \\
 {\small $^1$ Department of Physics, Shahid Rajaei University,
Lavizan, Tehran 16785-136, Iran
}\\
{\small $^2$ Departement Fysica, Universiteit Antwerpen,
Groenenborgerlaan 171, B-2020 Antwerpen,
 Belgium} }
\begin{document}
\title{Nanoindentation of a circular sheet of bilayer graphene}
\maketitle
\begin{abstract}
Nanoindentation of bilayer graphene is studied using molecular
dynamics simulations. We compared our simulation results with those
from elasticity theory as based on the nonlinear F\"{o}ppl-Hencky
equations with rigid boundary condition. The force deflection values
of bilayer graphene are compered to those of monolayer graphene.
Young's modulus of bilayer graphene is estimated to be 0.8 TPa which
is close to the value for graphite. Moreover, an almost flat bilayer
membrane at low temperature under central load has a 14$\%$ smaller
Young's modulus as compared to the one at room temperature.
\end{abstract}

\section{Introduction}
Graphene is an almost flat one-atom-thick layer of carbon atoms that
are densely packed in a honeycomb crystal lattice. Most of previous
studies concerned the electronic properties of graphene, but these
two dimensional graphene membranes have also exceptional mechanical
properties~\cite{geim,Giem2008,novoselov}. What makes graphene so
exceptional is that it stays strong and stiff even up to a single
atomic layer which differs from most materials whose mechanical
properties substantially deteriorate as they are made thinner. These
mechanical properties make graphene suitable for applications in e.
g. pressure sensing. A nano-membrane made of graphene, the ultimate
down-scaling limit of a nanoelectromechanical system, acts as a
sensor that is predicted to detect ultra-small masses and being at
the same time extremely robust against long-term wear.

 Recently, a nonlinear behavior of
stress-strain response of monolayer graphene was studied by Lee {\it
et al}~\cite{Changgu} using atomic force microscopy (AFM). They
measured Young's modulus and found that graphene is a strong
material like diamond.
 Furthermore, linear force-displacement curves
were measured in Refs.~\cite{frank,scott,scott2}. They obtained an
effective spring constant for a micron size monolayer graphene sheet
equal to 0.2~N/m~\cite{scott2} and for a suspended micron size sheet
of multilayer graphene with thicknesses in the range 2-8~nm and
found values in the range 1.0-5.0~N/m~\cite{frank}. Moreover,
monolayer graphene has a negative thermal expansion up to 900~K and
its Young's modulus increases with temperature up to
900~K~\cite{fasolino}. These anomalous properties are a consequence
of the strong anharmonicity in graphene. Most of these properties
can be explained by traditional elasticity theory and statistical
physics~\cite{frank}.

 The aim of the present
paper is to investigate if two van der Waals coupled graphene
layers, also called bilayer graphene, has similar improved
mechanical properties as monolayer graphene.

There are three common theoretical methods to study the indentation
of a graphene sheet, finite element method, continuum mechanics and
molecular mechanics simulations~\cite{sergry,hemmasizadeh,xu}. These
methods lead to different results for the force deflection curve and
for the graphene deformation profiles~\cite{xu}. The discrepancy
originates from the predominant bond stretching mode predicted by
the molecular mechanics model and a bending to stretching transition
process under increasing deflection predicted by continuum
mechanics.
 Nowadays various properties
are measured on micron size circular graphene~\cite{Changgu}, while
computational studies on nanoindentation of graphene typically are
limited to nano size systems. In a typical experimental setup, the
graphene sheet is indented by a tip from an atomic force microscope
(AFM).

An important issue in simulating  nanoindentation of graphene is the
type of interatomic potentials between the carbon atoms in the sheet
and between the carbon atoms and the atoms of the tip. In recent
studies on nanoindentation of graphene several models were used.
\emph{Xu et al}~\cite{xu} used 1392 atoms (a circular graphene sheet
with radius 3.25~nm)
   in their atomic simulation and obtained the elasticity response of such a
   nano circular sheet of graphene under central load.
    The nanoindentation was realized by moving down the center of the sheet by hand.
    On the other hand, Hemmasizadeh \emph{et
al}~\cite{hemmasizadeh} compared the results for the force
deflection curve obtained from a continuum elasticity
model~\cite{tim} and finite element method calculations for two
circular and hexagonal continuum plates of size 11~nm and showed
that in the large deflection regime there are important differences
between these two approaches. This difference roots in the effect of
exerting a concentrated force so that the difference reduced to half
when a spherical indenter with radius 10~\AA~ was used in the finite
element method calculation.

Medyanik \emph{et al}~\cite{sergry} studied the nanoindentation of a
hexagonal graphene sheet and multi-layered graphite sheets by
employing a quasi-static formulation of the method of multi-scale
boundary conditions. They showed that their method of multi-scaling
gives good results in comparison to full domain atomistic
simulations based on molecular mechanics simulations. They used
repulsive interactions to model the spherical indenter. The number
of atoms in the reduced domain was 2646 for single circular
graphene. Depending on the size of the reduced domain, their results
for multilayer graphene showed an increased force deflection
dependence as compared to monolayer graphene.

In this paper we study the mechanical properties of bilayer graphene
for three different sizes and compare the mechanical response of
bilayer graphene to the one of monolayer graphene. The Young's
modulus of bilayer graphene is estimated by using the predictions
from the theory of elasticity for a loaded plate in the large
deflection regime. Furthermore, the temperature dependence of the
nanoindentation of bilayer graphene is investigated.

This paper is organized as follows. In Sec.\,2  we will introduce
the atomistic model, the simulation method and the the results from
elasticity theory. Section\,3 contains the numerical results and in
Sec.\,4 we will present our conclusion.

\section{Theory and model}
We have used classical atomistic molecular dynamics simulation (MD)
to simulate the nanoindentation of a suspended  sheet of bilayer
graphene. The number of carbon atoms, i.e $N_b$, varies from 35,688
to 152,308 which are equivalent to surfaces with radius $R$=12~nm
and $R$=25~nm, respectively. A rigidly clamped boundary condition
was imposed. In our simulations, the indenter consists of $N_t$=371
atoms in a fcc structure (lattice constant
 equal to 3.92~\AA) and it is assumed rigid during our simulation.  The shape of the indenter
was chosen as a square-based pyramid. The area of the bottom surface
(square) of the tip is 2.02~$nm^2$ and the top one has an area of
10.24~$nm^2$. Initially the coordinates of all atoms in each
layer are put in a flat surface of a honey comb lattice with nearest
neighbor distance equal to 0.142 nm while the upper layer is shifted
along armchair direction by 0.142 nm above the bottom layer. The
separation between the two layers is 3.5~\AA. The initial
velocities in each direction, were extracted from a Maxwell-Boltzman
distribution for the given temperature. We simulated the system at
room temperature 300~K and 20~K by employing a Nos'e-Hoover
thermostat.
 The Brenner's bond-order potential~\cite{brenner, neek-amal} was
 used for the carbon-carbon interaction and a Lennard-Jones potential
 (LJ) $U(r)=4\varepsilon\{(\sigma/r)^{12}-(\sigma/r)^{6}\}$ for the indenter-graphene
interaction and the interaction between the two graphene layers. In
the LJ potential, $\sigma$ is the distance at which the potential is
zero and $\epsilon$ is the depth of the potential well.  The van
der-Waals interaction between the two graphene layers was modeled by
a LJ interaction with $\varepsilon_{C}=2.84$~meV and
$\sigma_{C}=$3.4~\AA. For the interaction between the tip and
graphene, we used the LJ parameters for Pt atoms with
$\varepsilon_{Pt}$=68.3~meV and
$\sigma_{Pt}=$2.54~\AA~\cite{erkoc2001}.  For a two-component
system, as studied here, the parameters for the mixed
interaction between the two type of atoms can be estimated by the
simple average $\sigma_{C-Pt}=(\sigma_{C}+\sigma_{Pt})/2$ and
$\varepsilon_{C-Pt}=\sqrt{\varepsilon_{C}\cdot\varepsilon_{Pt}}$
suggested by Steel {\it et al}~\cite{steel}. To save computational
time, we truncated the LJ potential at the cut-off distance of $r_c$
= 3.5 $\sigma$. Note that the LJ potential is a simple choice for
modeling the interaction between two layers~[18]. To obtain more
accurate results one can use other potentials such as a Morse
potential or other force fields~\cite{erkoc2001}. At the
start of our simulation, the position of the lowest atoms of the tip
are located a few angstroms, i.e. $\approx$3.4~\AA~ above the upper
graphene layer. After equilibrating the system during
$50000$ time steps, the indenter is pushed down slowly with
$\delta$=0.2~\AA~ in a time span of $5000\Delta t$ which is
equivalent to a velocity of 8~m/s, where $\Delta t$=0.5 fs is the time
step in our simulation. To avoid unphysical effects  due to
the  time step, the indentation step -$\delta$- was chosen
to be small with respect to the force cut-off length -$r_c$- for the
interatomic potential (e.g. to prevent a sudden change in the force, etc).

The considered size of the system ($\sim nm$) is larger than the
deflection value ($\sim \AA$), and in addition the thickness of the
sheet is smaller than the amount of deflection. Therefore, nonlinear
elasticity theory~\cite{landau} for a circular flake in the large
deflection limit along the $z$ direction is applicable. The energetics of
such a circular flake in the limit of large deflection is considered
including both bending and stretching energies~\cite{landau}. The
condition of minimum energy for the flake yields the F\"{o}ppl
equation. The solution of the equation is obtained by using the
Hencky transformation. The governing equations in planar-polar
coordinates are as follows
\begin{eqnarray}\label{Eq1}
r\frac{d}{dr}[\frac{d}{r dr}(r^2 \sigma_{r})]&=&-\frac{t
E}{2}(\frac{dz}{dr})^2, \nonumber\\\sigma_{r}
\frac{dz}{dr}&=&-\frac{F}{2\pi r},
\end{eqnarray}
 where $r$ is the radial
position, $R$ is the radius of the circular plate as shown in
Fig.~\ref{figmodel}, $z(r)$ is the deflection at radial position
$r$, $t$ is the thickness of the plate, $E$ is Young's modulus,
$\sigma_{r}$ is the radial stress of the flake and $F$ is the
concentrated load on the flake. We are interested in the situation
of a rigidly clamped boundary condition or fixed boundary condition
in the absence of residual stress, i.e., $z=0$ at $r=0$. Expressions
given by Eq.~(\ref{Eq1}) are nonlinear equations, however they can
be solved analytically in the special case of fixed boundary
condition~\cite{cong}. In general, the solution is given
by~\cite{cong}
\begin{equation}\label{Ftheory}
F=\frac{\pi E t}{4R^2}\frac{1}{G(\nu)}z(r)^3~,
\end{equation}
where $G(\nu)$ is a complicated function of the Poisson ratio,
$\nu$. $G(\nu)$ has almost a linear dependence on the Poisson's
ratio of the desired system and varies from $0.9$ for $\nu\simeq0.0$
to $0.65$ for $\nu\simeq0.5$.

 $\zeta$ is taken
 as the deflection of the bilayer graphene  at $r=0$, i.e. $\zeta=z(0)$. It is the difference between the center of mass of the central point of indented bilayer
graphene and the first central non-indented equilibrium position of
the bilayer (at $r=0$)[19] (see Fig. 1). During indentation, because
the using of the LJ potential, the equilibrium distance between the lowest
atoms of the tip and the central point of the bilayer is of the
order of $\sigma_{C-Pt}$. Surprisingly, our computer simulations
confirm this behavior which we will discuss in the next section.
 The force-displacement curves have been measured recently by Lee {\it et al}~\cite{Changgu}
 and they showed that it can be approximated by a simple
polynomial function having a linear and a cubic term,
\begin{equation}\label{Fexp}
F=a \zeta+b \zeta^3.
\end{equation}
When the bending stiffness is negligible and the load is small the
force deflection can be approximated by the linear term while the
second term dominates for large deflection. This behavior was
studied both theoretically and experimentally in
Refs.~\cite{Changgu,tim,poisson}.

\section{Numerical results}

Figure~\ref{figmodel} gives a schematic view of the system showing
the indenter and the clamped circular bilayer graphene and a
snapshot of an atomistic indenter over a circular bilayer graphene
with clamped boundary conditions. The $z$-component of the forces
from  the bilayer graphene atoms on the indenter are calculated by
summing over the total reaction forces:
\begin{equation}\label{Fz}
F={\sum^{N_t}_{i=1}}{\sum^{N_b}_{j=1}}
{F^z}_{ij}=24\varepsilon{\sum^{2}_{k=1}}\{{\sum^{N_t}_{i=1}}{\sum^{N_b}_{j=1}}
(-1)^k{2^{k-1}(\frac{\sigma}{r_{ij}})^{6k}
\frac{z_{ij}}{{r_{ij}}^2}}\}.
\end{equation}
Often in molecular dynamics simulations one approximates the above
sums by including only the nearest neighbors in order to reduce the
number of interactions which is accurate in the case of short range
potentials. Regarding the cut-off distance ($r_c$), only
those bilayer atoms below the tip interact most strongly with the
tip atoms, while outside this region the interaction strength
decreases very fast. Therefore, in  practice the sum over $N_b$ can be
truncated and limited to the atoms below the tip. This is done by
employing a neighbor list in our molecular dynamics simulation. The
upper (bottom) layer atoms of the bilayer, below the tip region,
interact with maximally $N_t\sim$300 (250) tip atoms for small
deflections and $N_t\sim$320 (265) for large deflections. Moreover,
the first term in Eq. (\ref{Fz}), i.e. $k$=1 in the parentheses, is the
derivative of the attractive part of the LJ potential and the second
term ($k$=2) is related to the derivative of the repulsive term.
We will compare our obtained Young's modulus
with those measured in experiments on graphene and
graphite~\cite{Changgu,Blakslee} and compare our simulated force
deflection curves of the clamped circular bilayer graphene with
those obtained for monolayer graphene~\cite{neek}. Circular red data in Fig.~\ref{figfvsz} show the variation of the
applied load at $r=R$ as a function of the deflection in the
$z$-direction. The fluctuation in the data are larger than those in our previous study on monolayer graphene. The reason is that in
bilayer graphene both layers vibrate (due to thermal fluctuations)
almost independently.

 The plotted force was obtained as follows. After each
5000 time steps the tip is pushed down with 0.2~\AA~ in order to
induce the deflection $\zeta$. During this time interval we let the
system equilibrate and in the last 1000 time steps of these
intervals we calculate $F$ and obtain the mean value of $F$. Thus in
a simulation with $10^6$ time steps we have 200 points in Fig.
{\ref{figfvsz}. The tip atoms are closer to the upper graphene layer
and the other layer is, on the scale of the tip interaction
potential, far from the tip. The latter layer is more free to
vibrate and it has a smaller interaction with the tip atoms.
Therefore, the contribution of the bottom layer to the summation of
Eq.~(\ref{Fz}) is less than the contribution of the other layer.
In our simulation we did not observe any defect formation
even for large deflections. Moreover, all the bilayer data of Figs.
{\ref{figfvsz}} and \ref{figflat} show  steps in the
force-deflection curves, that correspond to structural relaxation in
the process of indentation. The reason is that the number of
bilayer atoms which repel the tip are almost constant during the
indentation steps. The reason that the bilayer is a discrete sheet
and when it is indented by 2~\AA~the number of tip neighbors (from
bilayer atoms), which interact with the tip, are changed
non-continuously. Such steps can be smoothed by time averaging. As
we see the curve for R=12 nm is much smoother (than the two others
in Fig. 2), since it was averaged over six simulations with
different initial velocities.

The numerical results in Fig.~\ref{figfvsz} are fitted to the
expression given by Eq.~(\ref{Fexp}) and are shown by the solid
curves on the circular red data. The values for the fitting
parameters $a$ and $b$ are presented in Table 1. The results in
Fig.~\ref{figfvsz} show clearly the size dependence of
  the force-displacement curves and its difference
from similar results for monolayer graphene which are shown by
triangular green data in each panel of Fig.~\ref{figfvsz}. The
behavior of the tip-graphene flake interaction even for a few
angstroms displacement can be understood from our simulation given
here. To obtain quantitative results, we start with
Eqs.~(\ref{Ftheory}) and (\ref{Fexp}) to describe the dependence of
parameter $b$ on the radius for only large deflections. It is easy
to obtain an analytical expression for the $b$
 value for a large circular sample~\cite{cong} which is given by
\begin{equation}
b\cong\frac{\pi E t}{4G(\nu)}\frac{1}{R^2}~. \label{b}
\end{equation}

 Note that the parameter -$a$- in Eq.~(\ref{Fexp}) is
equivalent to an effective spring constant for perpendicular
indentation of the bilayer graphene when one assumes the bilayer
graphene as an elastic membrane. Furthermore, as can be seen from
Fig. 2, for small deflection the forces are almost linear,  we used
a linear fit ($F_l(\zeta)=a'\zeta$) for small deflections
($\zeta\leq1$ nm of the order of $\sqrt{a/b}$) and for large
deflections ($\zeta>1$ nm) we used a cubic fit
($F_c(\zeta)=b'\zeta^3+C$). The obtained  values for the parameter
$a'$ and $b'$ are listed in Table 1. For R=12, 15 and 18 nm the
parameter $C$ is 11.3 , 8.0 and 6.5 nN, respectively, which are of
the order of $a$.

The fitted curves are also shown in Fig.~2 by dash-dotted
curves. Figure~3 shows the obtained $b$ and $b'$ values as a
function of the bilayer graphene circular size. The fitted parameter
$b'$ follows very well a $1/R^2$ function and the agreement is now
better than for the previous b-values.

For monolayer graphene, we fitted the function
$F=a_m\zeta+b_m\zeta^3$ to the results reported in our previous
study~[24]. The corresponding values for $a_m$ and $b_m$ are listed
in Table~1. The fitted curves are shown by solid curves. Analogous
 as for bilayer graphene we also used separate fits for small and large deflection: $F_l(\zeta)=a_m'\zeta$ and
$F_c(\zeta)=b_m'\zeta^3+C'$, respectively, which are presented in
Table 1. The corresponding fitted curves are shown by the dash-dotted
curves in Fig.~2. For R=12, 15 and 18 nm the parameter $C'$ is 8.1
, 3.7 and 3.9. nN, respectively, which are of the order of $a'$.

\begin{table}
\begin{tabular}{|c|c|c|c|c|}
 \hline
  R(nm) & $a$(N/m)&$b(10^{17}$N/m$^3$)&  $a'$(N/m)&$b'(10^{17}$N/m$^3$)\\
     \hline
  12 & 7.10$\pm$0.4& 20.20$\pm0.5$&7.15$\pm$0.9&$23.3\pm$0.6\\
  15 & 3.80$\pm$0.5& 15.60$\pm0.2$&7.17$\pm$1.3&16.77$\pm$0.8\\
  18 & 4.60 $\pm$0.5&10.90$\pm0.3$&3.19$\pm$0.9&11.99$\pm$1.1\\
  25& 5.50 $\pm$0.3&4.20 $\pm0.2$&5.30$\pm0.9$&6.04$\pm0.9$\\
\hline
  R(nm) &$a_m$(N/m) &$b_m(10^{17}N/m^3)$& $a'_m$(N/m)&$b'_m(10^{17}N/m^3)$\\
  \hline
  12 & 5.21$\pm0.08$&10.30$\pm0.07$&3.41$\pm0.1$&13.48$\pm0.1$ \\
  15 & 3.16$\pm0.06$&7.07$\pm0.01$&2.65$\pm0.3$&10.07$\pm0.3$  \\
  18 &3.06$\pm0.04$&4.46$\pm0.02$&1.86$\pm0.1$&5.30$\pm0.2$  \\
  25&2.53$\pm0.07$&2.70$\pm0.05$&1.50$\pm0.2$& 4.01$\pm0.2$  \\
  30&1.30$\pm0.09$&2.03$\pm0.06$&0.65$\pm0.1$& 2.30$\pm0.2$ \\
    \hline
\end{tabular}
  \centering
\caption{Values of $a$ and $b$ determined by fitting
Eq.~(\ref{Fexp}) to our simulated data given in Fig.~\ref{figfvsz}.
Values of $a'$ and $b'$ are determined by fitting separately the linear and cubic
equations with the small and large deflection regions, respectively.
The parameters $a_m$ and $b_m$ and also $a'_m$ and $b'_m$ for
monolayer graphene which we determined earlier \cite{neek} are also
shown.}\label{table1}
\end{table}

Furthermore, we fitted the values for $b'$ to Eq.~(\ref{b}) and
obtained $\pi E t/ G(\nu)\approx 4\times353.5$~N/m. This result is
obtained from the function G($\nu$)=-0.59$\nu$+0.94 obtained from
Ref.~\cite{cong} and using the value $\nu\approx0.25$ which is a
typical Poisson's ratio for monolayer graphene ~\cite{poisson}.
To obtain the Young modulus we need an appropriate value for the
thickness of bilayer graphene. One can estimate $t$ as the distance
between the two layers plus the thickness of a monolayer. For
monolayer graphene some experimentalists used the value 0.23~\AA~by
assuming Young's modulus of graphene to be the same as in
graphite~\cite{Giem2008}. Others use different values for the
thickness of monolayer graphene. Saitoh \emph{et al} estimated
0.874~\AA~which was obtained from the parameters of the Brenner
potential~\cite{Saitoh}. This is an independent and more accurate
value. Using this value the bilayer thickness becomes  4.5~\AA which leads to the Young modulus $0.8$~TPa which is
close to the values for graphite~(i.e. 1.02 $\pm$
0.03~TPa~\cite{Blakslee}) and those, found in recent experiments and
found in theoretical calculations, for graphene~(i.e. 1.0$\pm$
0.1~TPa~\cite{Changgu}).

In order to investigate the effects of having two instead of a
single graphene layer we calculated the ratio
$\frac{F_{bilayer}}{F_{monolayer}}$ which we show in
Fig.~\ref{figfoverf}. For small indentation the fluctuations in our
results are larger because both layers vibrate more freely, while in
the large deflection regime the elastic membrane energy is dominant.
Furthermore, because of the nature of the LJ potential, the
repulsion of bilayer graphene with the tip is more or less two times
larger than for single monolayer graphene (see Fig.
\ref{figfoverf}). As can be seen from Fig.~\ref{figfoverf}, for
small deflection where the Hooke's regime is dominant, we see a
large difference between the spring constant ($a$ parameters) of
bilayer graphene and
 monolayer graphene. In the linear regime in Fig.~\ref{figfoverf} we have
$\frac{F_{bilayer}}{F_{monolayer}}\approx\frac{a_b}{a_m}$, where
$a_b(=a)$ and $a_m$ are the effective spring constant of bilayer
 and monolayer graphene. Figure \ref{figfoverf} shows that
in the small $\zeta-$region we have on the average
$F_{bilayer}<2F_{monolayer}$ while in the large deflection regime
$F_{bilayer}\simeq2 F_{monolayer}$. Note that the ratio
$\frac{F_{bilayer}}{F_{monolayer}}$ in the large deflection limit is
not exactly 2, because the bottom layer of bilayer graphene is far
from the tip with respect to the upper layer so the number of
interacting atoms in bilayer graphene to the tip at fixed deflection
is not exactly twice those of monolayer graphene. Note that
for small deflection (when the sheet is not stressed)  shoulders
(with small upwards amplitudes) are seen  in both layers of the
bilayer (particularly in the bottom layer because it is more free)
around and
 below the tip which attract (they are close to the tip) the tip and cause negative forces even
for small positive deflections (as measured from below the tip).
This attraction dominates the repulsion term that is due to  the other
atoms below the tip. This unusual behavior is more clearly seen for
larger systems.

At low temperature ($T$=20~K) graphene has a smaller roughness and
behaves as a flat honeycomb lattice. One can compare the low
temperature data in Fig.~\ref{figflat} (an almost flat bilayer
graphene with $R$=12~nm) to those for room temperature. At low
temperature and for a fixed value of the displacement the forces are
smaller than those at room temperature. We fitted Eq.~(\ref{Fexp})
to the data and found $a_{l}=2.45~$N/m and
$b_{l}=18.5\times10^{17}N/m^3$. Using Eq.~(\ref{b}) yields
Young's modulus 0.69~TPa~for low temperature which is 14$\%$
smaller than Young's modulus for bilayer graphene at room
temperature. Such an increase of the Young's modulus with
temperatures is unusual but it is similar to what has been reported
recently by Zakharchenko~\emph{et al}~\cite{fasolino} for monolayer
graphene. This unusual behavior in both monolayer and bilayer
graphene is the consequence of strong anharmonicity in graphene.
\section{Conclusion}
In this study we showed that bilayer graphene like monolayer
graphene and other carbon nanostructures has an exceptional
stiffness. Young's modulus for bilayer graphene was calculated and
found to be 0.8~TPa. The force-displacement result could be fitted
to the function $F=a\zeta+b\zeta^3$. We found that for given
displacement the exerted force on bilayer graphene has to be about
twice the one on monolayer graphene. The force-deflection result is
found to be temperature dependent. At low temperature Young's
modulus is found to be 14$\%$ smaller than at room temperature.

\section{Acknowledgment}
We gratefully acknowledge comments from R. Asgari. M. Neek-Amal
would like to thank the Universiteit of Antwerpen
 for its hospitality where part of this work was performed.
 This work was supported by the Flemish science foundation (FWO-Vl) and the Belgium Science Policy~(IAP).

\pagebreak
\begin{figure}[ht]
\begin{center}
\includegraphics[width=0.7\linewidth]{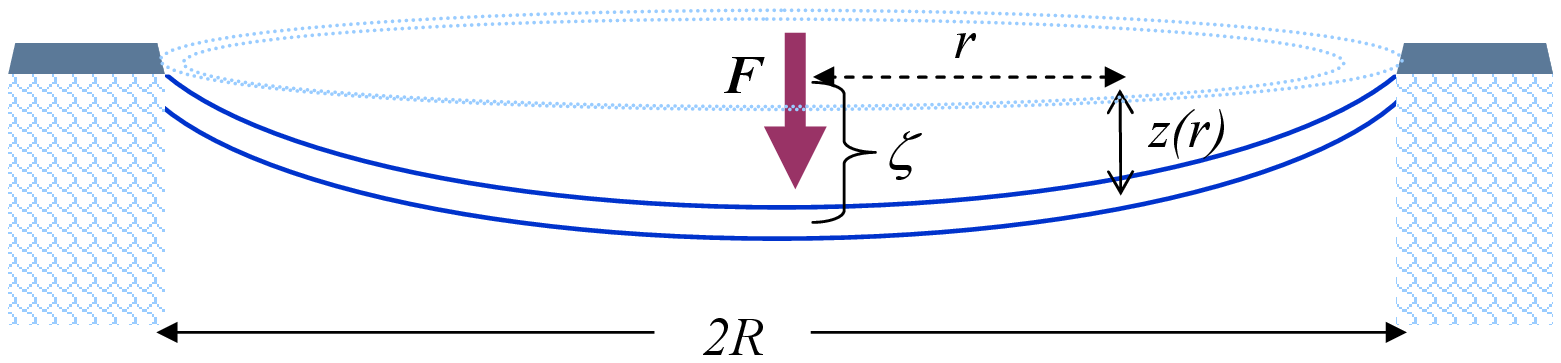}
\includegraphics[width=0.7\linewidth]{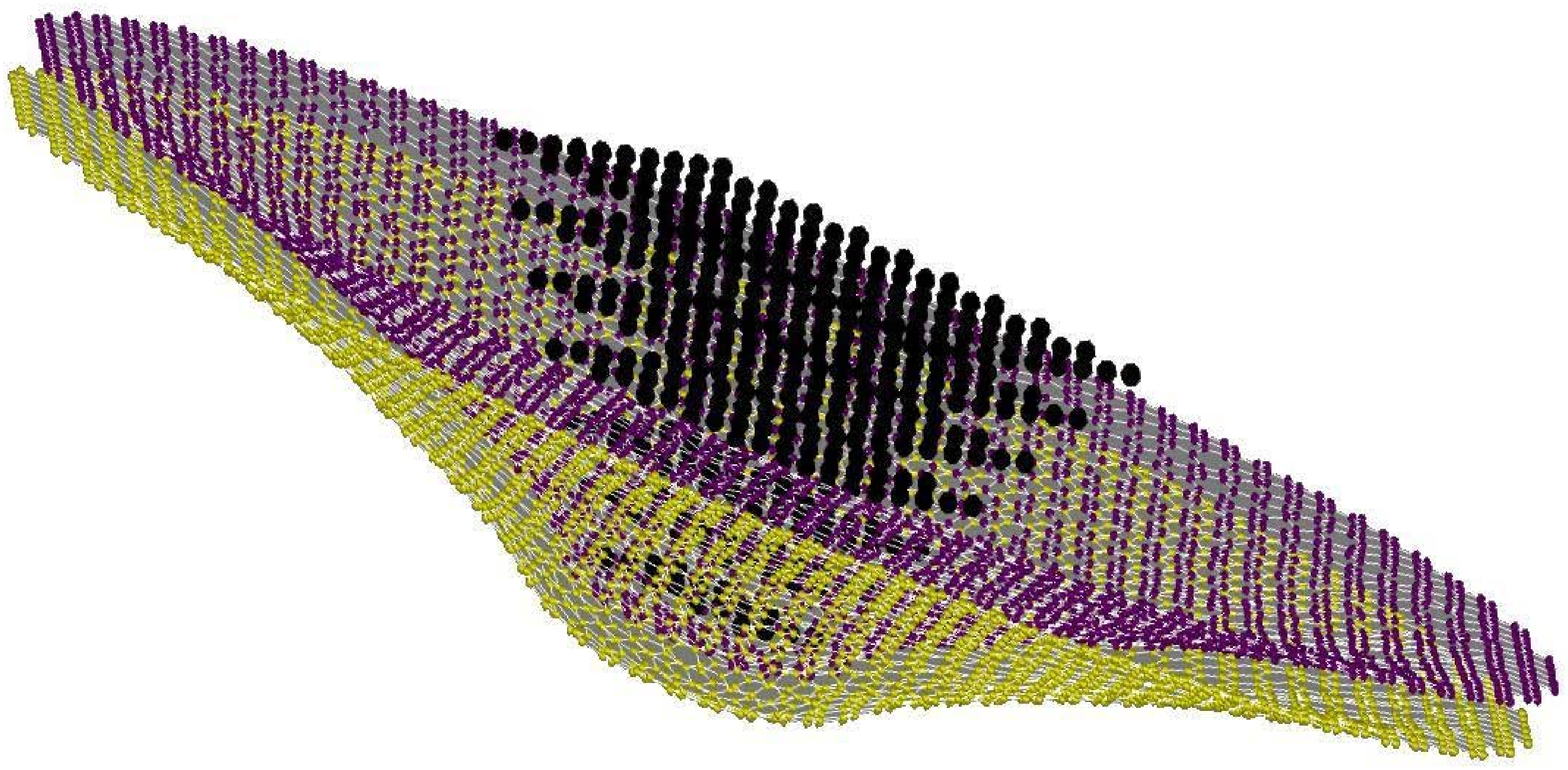}
\caption{(Color online) Top panel: Schematic model of the suspended
 bilayer graphene with rigidly clamped boundary conditions. Bottom panel: A
snapshot of an atomistic indenter over a circular bilayer graphene
with $R=$6~nm and clamped boundary conditions.\label{figmodel} }
\end{center}
\end{figure}

\begin{figure}[ht]
\begin{center}
\includegraphics[width=0.4\linewidth]{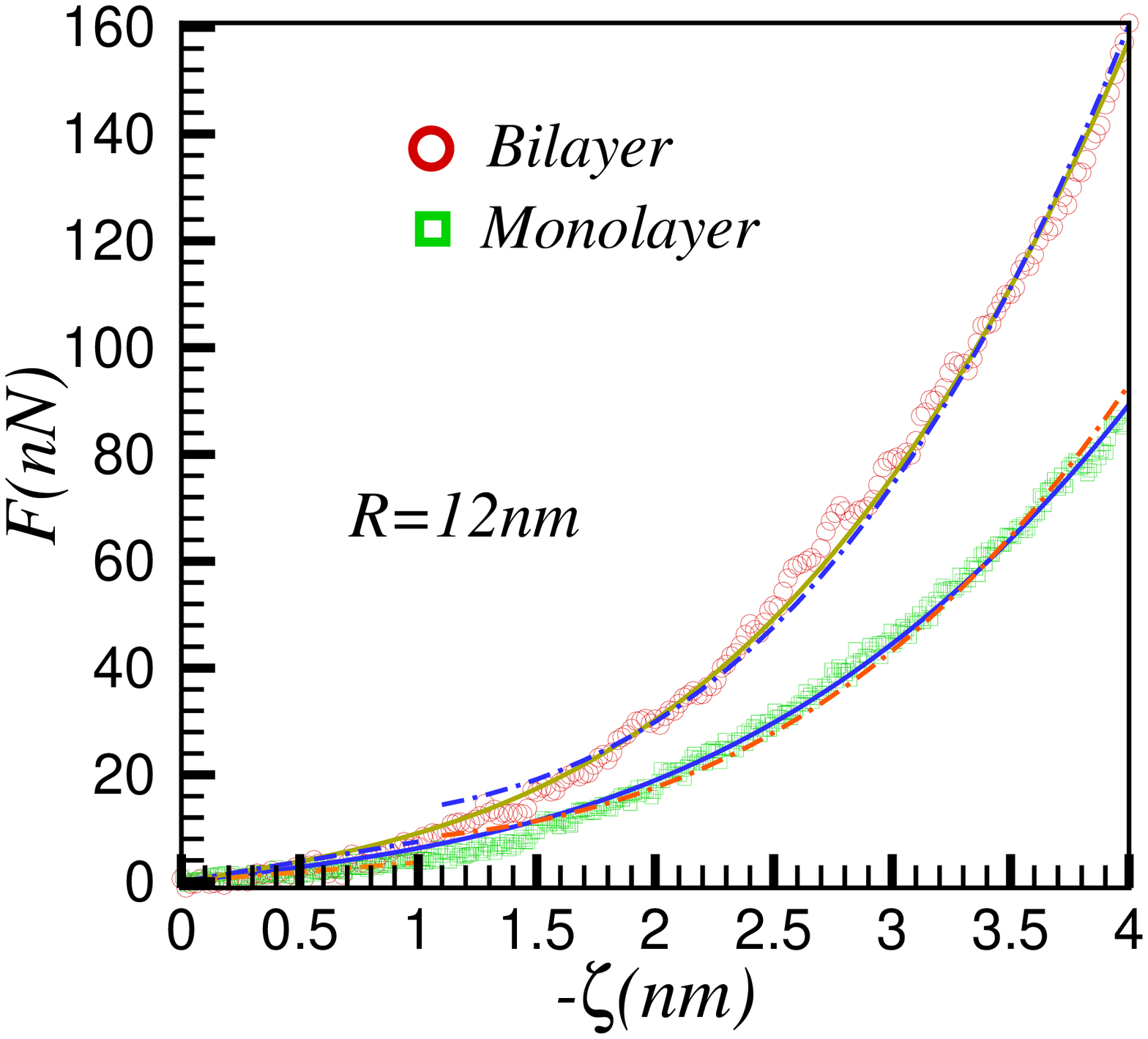}
\includegraphics[width=0.4\linewidth]{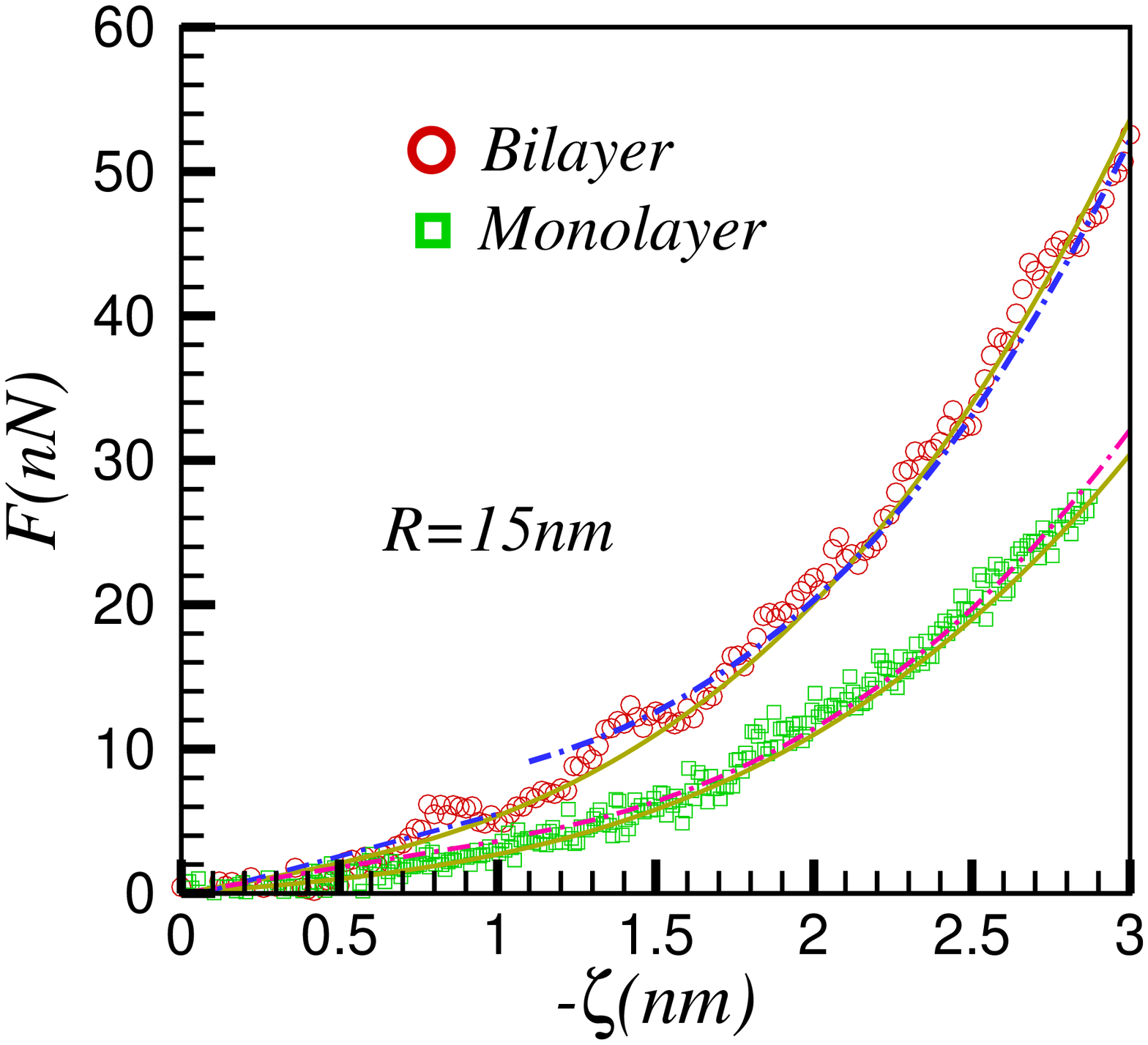}
\includegraphics[width=0.4\linewidth]{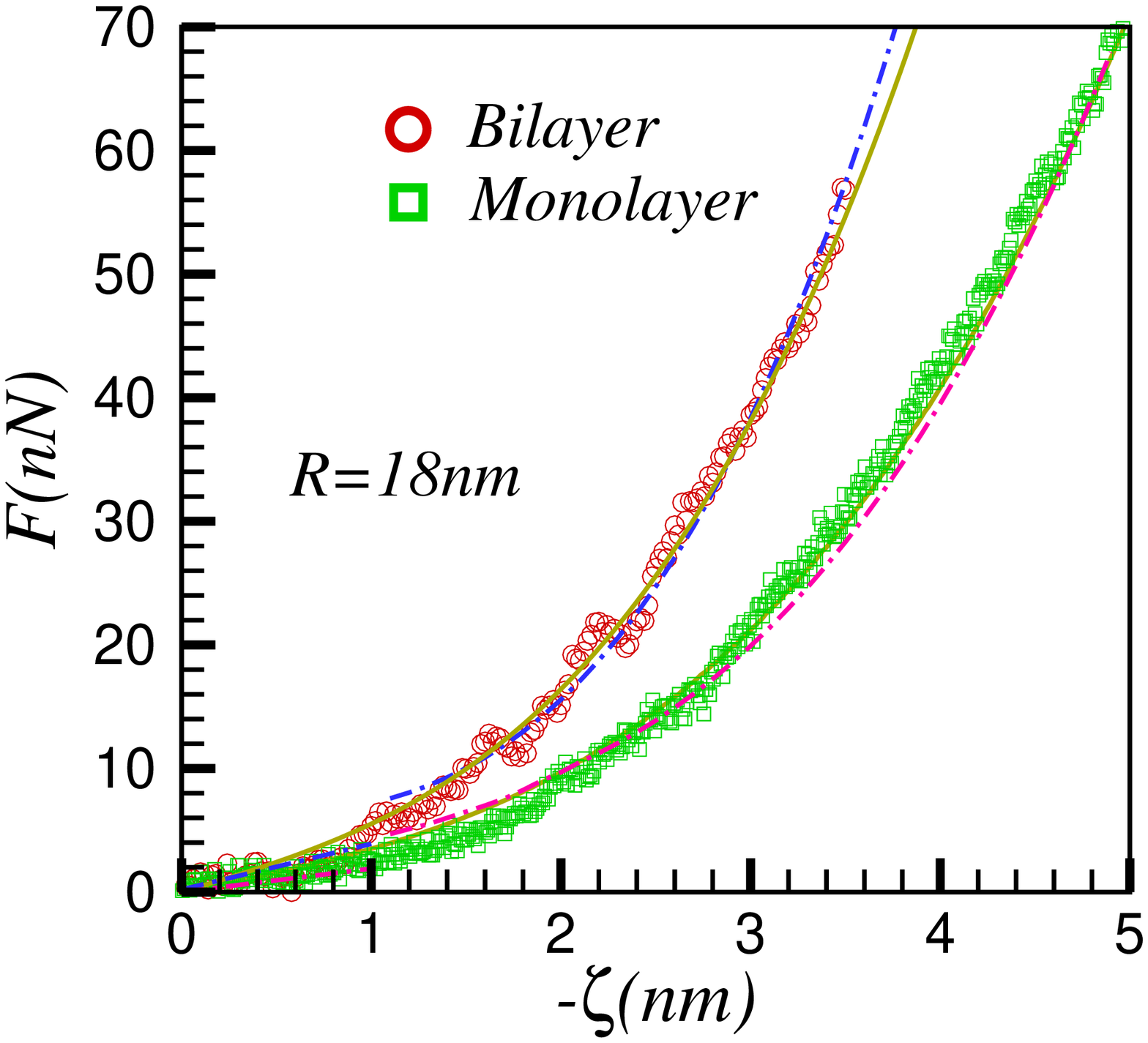}
\caption{(Color online) Load force as a function of displacement for
three different radii. The analytical expression Eq.~(\ref{Fexp}) is
fitted and shown by the solid curves. \textbf{The dash-dotted curves
refer to two separate fits for small and large deflections.} For
comparative purposes we show also the results of monolayer graphene
and corresponding fits~\cite{neek} . \label{figfvsz} }
\end{center}
\end{figure}

\begin{figure}[ht]
\begin{center}
\includegraphics[width=0.6\linewidth]{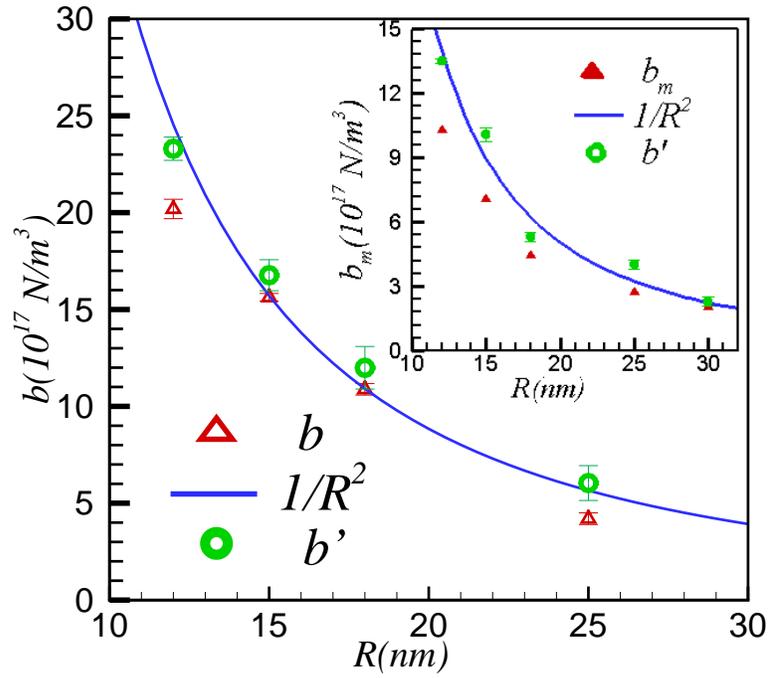}
\caption{(Color online) The parameters $b$ and $b'$ as a function of
the circular size of bilayer graphene. The inset shows the parameters
$b_m$ and $b'_m$  as a function of the circular size of
monolayer graphene. In both cases the solid curves show a $1/R^2$
function fitted to the $b'$ and $b'_m$ data. For the definition of
$b,~b_m$ and $b',~b'_m$ we refer to the text.~(i.e. Eqs.
(\ref{Fexp}) and (\ref{b}), respectively) .\label{figbvsR}}
\end{center}
\end{figure}

\begin{figure}[ht]
\begin{center}
\includegraphics[width=0.6\linewidth]{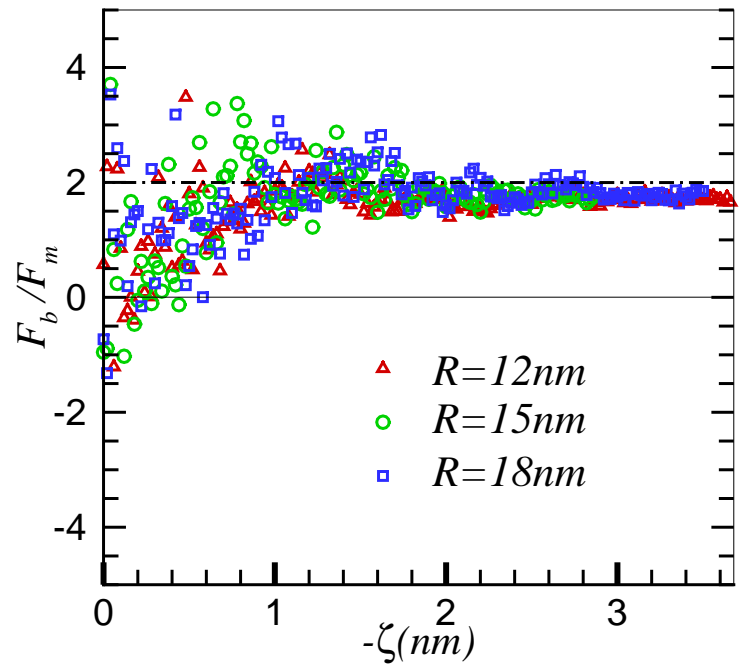}
\caption{(Color online) The ratio between the forces in loaded
bilayer graphene and loaded monolayer graphene for three values of
the size of the membrane.\label{figfoverf} }
\end{center}
\end{figure}

\begin{figure}[ht]
\begin{center}
\includegraphics[width=0.6\linewidth]{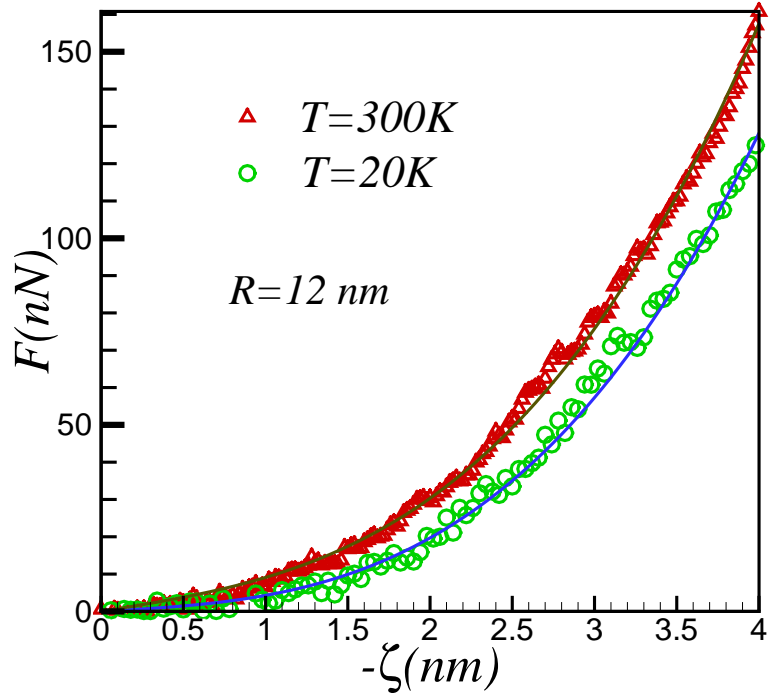}
\caption{(Color online) Comparison between the force deflection
response for rough (i.e. T=300K) and almost flat (i.e. T=20K) bilayer graphene for a membrane
with radius $R$=12~nm. Solid curves are the fitted results using
 Eq.~(\ref{Fexp}) with fitting parameters $a=7.1$~N/m and $b=20.20\times10^{17}~N/m^3$ for $T$=300~K
 and also $a_{l}=2.45~$N/m and $b_{l}=18.5\times10^{17}N/m^3$ for $T$=20~K.\label{figflat}}
\end{center}
\end{figure}

\end{document}